\documentclass[a4paper, 11pt]{article}
\usepackage{amsmath}
\usepackage{graphicx}
\usepackage{latexsym}

\def\b{\begin{eqnarray}}
\def\e{\end{eqnarray}}
\def\n{\noindent}

\newtheorem{lemma}{Lemma}

\begin{document}

\begin{center}

{\LARGE\textbf{Extended Camassa-Holm Hierarchy and Conserved
Quantities
\\}} \vspace {10mm} \vspace{1mm} \noindent

%Running title:   {\textbf{Extended Camassa-Holm Hierarchy\\}}
%\vspace {10mm} \vspace{1mm} \noindent

{\large Rossen I. Ivanov}\footnote{On leave from the Institute for
Nuclear Research and Nuclear Energy, Bulgarian Academy of Sciences,
Sofia, Bulgaria.} \vskip1cm \n \hskip-.3cm
\begin{tabular}{c}
\hskip-1cm $\phantom{R^R}${School of Mathematics, Trinity College,}
\\ {Dublin 2, Ireland} \\ \vspace {10mm}  \\Tel:  + 353 - 1 - 608 2898; \\Fax:  + 353 - 1- 608
2282; E-mail: ivanovr@tcd.ie \\
\\
\hskip-.8cm
\end{tabular}
\vskip1cm
\end{center}

\begin{abstract}
\n An extension of the Camassa-Holm hierarchy is constructed in this
letter. The conserved quantities of the hierarchy are studied and a
recurrent formula for the integrals of motion is derived.

\vspace {10mm}



\end{abstract}

\newpage

\section{Introduction}

The Camassa-Holm equation (CH) \cite{CH93}
\begin{equation}\label{eq1}
 u_{t}-u_{xxt}+2\omega u_{x}+3uu_{x}-2u_{x}u_{xx}-uu_{xxx}=0,
\end{equation}
where $\omega$ is a real constant parameter, describes the
unidirectional propagation of shallow water waves over a flat bottom
\cite{CH93, J02}. It firstly appeared in \cite{FF81} as an equation
with a bi-hamiltonian structure. CH is a completely integrable
equation \cite{BBS98, FOR96, CM99, C01, L02, GH03}, describing
permanent and breaking waves \cite{CE98, C00}. Its solitary waves
are stable solitons if $\omega
> 0$ \cite{BBS99, CS00, CS02, J03} or peakons if $\omega = 0$
\cite{CH93,LOR99}. CH arises also as an equation of the geodesic
flow for the $H^{1}$ metrics on the Bott-Virasoro group \cite{M98,
CK03, CKKT04}. The bi-Hamiltonian form of (\ref{eq1}) is \cite{CH93,
FF81, FOR96}

\begin{equation}\label{eq2}
m_{t}=-(\partial-\partial^{3})\frac{\delta H_{2}[m]}{\delta
m}=-(2\omega \partial +m\partial+\partial m)\frac{\delta
H_{1}[m]}{\delta m},
\end{equation}

\n where \b\label{eq4a} m = u-u_{xx}, \e \n and the Hamiltonians are
\b \label{eq2a} H_{1}[m]&=&\frac{1}{2}\int m u \text{d}x,
 \\\label{eq2b}
H_{2}[m]&=&\frac{1}{2}\int(u^{3}+uu_{x}^{2}+2\omega u^{2})\text{d}x.
\e

\n The integration is from $-\infty$ to $\infty$ in the case of
Schwartz class functions, and over one period in the periodic case.

In \cite{FS99} it is shown that CH has an infinite number of local
conserved quantities. A scheme for computation of the conservation
laws is proposed in \cite{R02,L05,CL05}. In this contribution we
present a scheme, providing an explicit recurrent formula for the
infinite sequence of independent integrals of motion for a chain of
CH type equations.

The equation (\ref{eq1}) admits a Lax pair \cite{CH93}

\b \label{eq3} \Psi_{xx}&=&\Big(\frac{1}{4}+\zeta
(m+\omega)\Big)\Psi,
 \\\label{eq4}
\Psi_{t}&=&\Big(\frac{1}{2\zeta
}-u\Big)\Psi_{x}+\frac{u_{x}}{2}\Psi. \e

\n Recently, various multi-component generalizations of the CH
equation are under intense investigation, e.g.
\cite{GH03,HQ03,CLZ05,EP05,KZ00} and probably some others. In this
work we also present a construction for a multi-component system,
which admits reduction to CH equation.

\n To this end, instead of the Lax pair (\ref{eq3}), (\ref{eq4}), we
consider a more general one, leading to a hierarchy of Camassa-Holm
type:

\b \label{L1} \Psi_{xx}&=&Q(x,\lambda)\Psi,
 \\\label{L2}
\Psi_{t}&=&-U(x,\lambda)\Psi_{x}+\frac{1}{2}U_x(x,\lambda)\Psi, \e

\n where

\b \label{L3} Q(x,\lambda)&=&\lambda^n q_n(x)+\lambda^{n-1}
q_{n-1}(x)+\ldots+\lambda q_1(x)+\frac{1}{4},
 \\\label{L4}
U(x,\lambda)&=&u_0(x)+\frac{u_1(x)}{\lambda}+\ldots
\frac{u_k(x)}{\lambda^k}. \e

\n The compatibility condition of (\ref{L1}), (\ref{L2}) gives the
equation

\b \label{L5} Q_t=\frac{1}{2} U_{xxx}-2U_x Q-UQ_x,\e

\n which, due to (\ref{L3}), (\ref{L4}), is equivalent to a chain of
$n$ evolution equations with $k+1$ differential constraints for the
$n+k+1$ variables $q_1$, $q_2$, $\ldots$, $q_n$, $u_0$, $u_1$,
$\ldots$, $u_k$ ($n$ and $k$ are arbitrary natural numbers, i.e.
positive integers):

\b  q_{n-r,t}&=&-\sum_{s=\max(0,r-k)} ^{r}(2u_{r-s,x}q_{n-s}+u_{r-s}q_{n-s,x}),\qquad r=0,1,\ldots,n-1, \nonumber \\
0&=&\frac{1}{2}(u_{r,xxx}-u_{r,x})-\sum_{s=1} ^{\min(n,k-r)}(2u_{r+s,x}q_{s}+u_{r+s}q_{s,x}), \nonumber \\
&\phantom{=}& \phantom{****************************}  r=0,1,\ldots,k-1, \nonumber\\
 0&=&\frac{1}{2}(u_{k,xxx}-u_{k,x}).\label{L6}  \e

%In the case $n>k$, the system is:

%\b  q_{n-r,t}&=&-\sum_{s=0} ^{r}(2u_{r-s,x}q_{n-s}+u_{r-s}q_{n-s,x}),\qquad r=0,1,\ldots,k, \nonumber \\
%q_{n-k-r,t}&=&-\sum_{s=0} ^{k}(2u_{k-s,x}q_{n-s-r}+u_{k-s}q_{n-s-r,x}), \nonumber \\
%&\phantom{=}& \phantom{*******************} r=1,\ldots,n-k-1, \nonumber\\
%0&=&\frac{1}{2}(u_{r,xxx}-u_{r,x})-\sum_{s=1} ^{k-r}(2u_{r+s,x}q_{s}+u_{r+s}q_{s,x}), \nonumber \\
%&\phantom{=}& \phantom{*********************}  r=0,1,\ldots,k-1, \nonumber\\
%0&=&\frac{1}{2}(u_{k,xxx}-u_{k,x}).\label{LL6}  \e

 The system (\ref{L6}) is similar to the hydrodynamic chain, studied
 in a series of papers \cite{MA02,MA03,MA04,Sh05}, and to other CH generalizations \cite{GH03,HQ03,EP05,KZ00,CGP97}.
 Let us now consider the following examples.

 {\it Example 1}: $k=n=2$.

 The choice $u_2=-1/2$ solves automatically one of the constraints.
 The other two differential constraints can be easily integrated,
 giving

\b \label{L7} q_{1}&=&u_1-u_{1,xx}+\omega_1,\\
q_2&=&u_0-u_{0,xx}+3u_1^2-u_{1,x}^2-2u_1u_{1,xx}+4\omega_1
u_1+\omega_2,\label{L8}\e

\n where $\omega_{1,2}$ are arbitrary constants. The system of
equations for $u_0$, $u_1$ is

\b  q_{2,t}\!\!&+&\!\!2u_{0,x}q_2+u_0q_{2,x}=0, \label{L9} \\
q_{1,t}\!\!&+&\!\!2u_{0,x}q_{1}+u_0q_{1,x}+2u_{1,x}q_{2}+u_1q_{2,x}=0.
\label{L10} \e

{\it Example 2}: $k=1$, $n=2$.

This system was studied in more details in \cite{CLZ05}. In the
notations $u_0\equiv u$, $q_1\equiv -q$ and $q_2 \equiv \rho^2$, and
with the choice $u_1=-1/2$, the system can be written in the form

\b q_t\!\!&+&\!\!uq_x+2q u_x -\rho \rho_x=0,\label{LL}\\
\rho_t\!\!&+&\!\!(u\rho)_x=0,\label{LQ}\e

\n where $q=u-u_{xx}+\omega $ and $\omega$ is an arbitrary constant.

{\it Example 3}: CH  equation

CH can be considered as a reduction from the system (\ref{L7}) --
(\ref{L10}). Indeed, one can obtain an integrable reduction of
(\ref{L7}) -- (\ref{L10}) by taking $u_1=\omega_1=0$. Then $q_1=0$,
$q_2=u_0-u_{0,xx}+\omega_2$ and (\ref{L9}) is exactly the CH
equation (\ref{eq1}) with $u\equiv u_0$ and $\omega \equiv
\omega_2$. Equation (\ref{L10}) is trivially satisfied. Thus, the CH
corresponds to a Lax pair with

\b \label{L11} Q(x,\lambda)=\lambda^2 q_2(x)+\frac{1}{4},\qquad
U(x,\lambda)=u_0(x)-\frac{1}{2\lambda^2}. \e

\n It is not difficult to recover the Lax pair (\ref{eq3}) --
(\ref{eq4}) by identifying $\zeta=\lambda^2$.

CH can also be obtained as a reduction from (\ref{LL}) -- (\ref{LQ})
by setting $\rho=0$.

\section{Generating Function for the Integrals of Motion}
Introducing

\b \label{eq40} p= \frac{\Psi_x}{\Psi}, \e \n from (\ref{L1}) we
obtain (cf. \cite{DHH02,G85,R02})

\b \label{eq10} p_x +p^{2}=Q(x,\lambda). \e

\n Then, from (\ref{L2}), (\ref{eq40}) and (\ref{eq10}) the
following conservation law follows:

\b \label{eq50} p_t= \Big( \frac{1}{2}U_x-pU\Big)_{x}. \e \n
Therefore $p(x,\lambda)$, regarded as a solution of (\ref{eq10}), is
the density of the generating function of the conservation laws. We
notice that the densities are determined up to a constant. Indeed,
if $\alpha(\lambda)$ is an arbitrary function of $\lambda$, the
quantity $P=p+\alpha$ is also a generating function, since it
satisfies \b \label{eq50a} P_t= \Big( \frac{1}{2}U_x-PU+\alpha
U\Big)_{x}. \e We can use this freedom to fix the convergency
properties of the integrals representing the conserved quantities.

Clearly, $p$ is related to the scattering matrix \cite{C01} when the
Schwartz class of solutions is considered, or to the monodromy
matrix in the periodic case \cite{G85}.

Now it is evident that (\ref{eq50}) or (\ref{eq50a}) represent a
parameter-dependent conservation law, which is equivalent to a
sequence of infinitely many conservation laws.

Indeed, since $\lambda$ is an arbitrary (spectral) parameter, one
can expand the solution $p(x,\lambda)$ of (\ref{eq10}) about
$\lambda=\infty$. Let us suppose for simplicity that $n=2a$ is an
even number. (The case when $n$ is odd is handled in a similar
manner, e.g. by introducing a new spectral parameter $\zeta$,  such
that $\lambda=\zeta^2$, see the CH case \cite{R02,CL05}, where
$n=1$). Then

\b \label{eqA} p(x,\lambda)=p_a\lambda^a+\ldots+p_1\lambda+p_0+
+\sum _{s=1}^{\infty} \frac{p_{-s}}{\lambda^{s}}. \e

\n There are finitely many positive powers in $\lambda$ due to their
presence in the RHS of (\ref{eq10}). From (\ref{eq50}) and
(\ref{eqA}) the following infinite sequence of conservation laws
follows (it is assumed $u_r\equiv0$ for $r>k$):

\b p_{a-r,t}&=&-\Big(\sum_{s=0}^{r} u_{r-s}p_{a-s}\Big )_{x}, \qquad
r=0,1,\ldots,a-1, \nonumber\\
p_{-r,t}&=&\Big(\frac{1}{2}u_{r,x}-\sum_{s=0}^{a+r}u_{a+r-s}p_{a-s}\Big
)_{x}, \qquad r=0,1,\ldots\phantom{*}. \label{eqDF1} \e

Let us now expand the solution $p(x,\lambda)$ of (\ref{eq10}) about
$\lambda=0$:

\b \label{eqB} p(x,\lambda)=p_0+\sum _{s=1}^{\infty}
p_{s}\lambda^{s}. \e

\n Note that the expansion coefficients $p_s$ in (\ref{eqB}) are not
the same as those in (\ref{eqA}). From (\ref{eq50}) and (\ref{eqB})
another infinite sequence of conservation laws follows:

\b p_{r,t}&=&-\Big(\sum_{s=0}^{k} u_{s}p_{s+r}\Big )_{x} \qquad
r=1,2,\ldots, \nonumber\\
p_{0,t}&=&\Big(\frac{1}{2}u_{0,x}-\sum_{s=0}^{k} u_{s}p_{s}\Big
)_{x}, \qquad \label{eqDF2}\e

\n and in addition, some of the constraints can be rewritten in the
form:

\b \Big(\frac{1}{2}u_{r,x}-\sum_{s=0}^{k-r} u_{r+s}p_{s}\Big )_{x}
=0,\qquad r=1,2,\ldots,k. \label{eqC} \e

In general, there are two different solutions of (\ref{eq10}).
However, these solutions do not represent independent integral
densities of conserved quantities, due to the following result,
showing that each of the solutions can be expressed linearly through
the other one, up to an exact derivative:

\begin{lemma}\label{lem} If $p^{+}(x,\lambda)$ and $p^{-}(x,\lambda)$ are two
different solutions of (\ref{eq10}), then there exists a function
$f(x,\lambda)$, such that

\b\label{eqL1}p^{+}(x,\lambda)=-p^{-}(x,\lambda)+f_x(x,\lambda). \e
\end{lemma}

{\bf Proof}. Writing (\ref{eq10}) for $p^{+}(x,\lambda)$ and
$p^{-}(x,\lambda)$ and subtracting the two equations we obtain
immediately $f(x,\lambda)=-\ln[p^{+}(x,\lambda)-p^{-}(x,\lambda)]$.

\section{Computation of the Conserved Quantities}

In order to find the integral densities, one needs to compute
explicitly the expansion coefficients in (\ref{eqA}), (\ref{eqB}).
In this Section we will illustrate the method at the system from
Example 1. (The conservation laws for the CH, Example 3, follow
immediately via the described reduction.) Clearly, one can apply an
analogous procedure to any particular case of the system (\ref{L6}).

Before going into the actual computations, the following observation
is in order. Exactly as in \cite{C01,C05}, using only (\ref{L9}) one
can prove that  $q_2(x,t)$  does not change sign if $q_2(x,0)$ does
not. The idea of proof is as follows. Consider the diffeomorphism of
the line $\varphi(x,t)$, such that

\b \label{eq51} \varphi_t=u_0(\varphi(x,t),t),\qquad \varphi(x,0)=x.
\e

\n The solution of (\ref{eq51}) is unique and represents an
increasing diffeomorphism ($\varphi_x>0$ for all $t$) of
$\textsf{R}\ni x$ \cite{C05}. Then, using (\ref{L9}) and
(\ref{eq51}), one can check that \b \label{eq52}
q_2(\varphi(x,t),t)\varphi_x^2(x,t)=q_2(x,0) \e and the claim easily
follows. Then, for simplicity, in order to make sense of the
expressions like $\sqrt{q_2(x,t)}$, $1/\sqrt{q_2(x,t)}$, etc; we
assume that the initial data is such that $q_2(x,0)$ does not change
sign, i.e. $q_2(x,0)>0$.

Also, in the case when $u_0$, $u_1$, ... are Schwartz class
functions, it may happen that the integral density at
$x\rightarrow\pm\infty$ is not zero but a constant. Then, in order
to make sense of the integral, the density obviously should be
reduced by the same constant, cf. (\ref{eq50a}).

The equation (\ref{eq10}) is

\b \label{eq14} p_x +p^{2}=\frac{1}{4}+ \lambda q_1+\lambda^2q_2, \e
and admits a solution of the form (\ref{eqA}): \b \label{eq15}
p&=&p_{1}\lambda +p_{0}+\sum _{s=1}^{\infty}
\frac{p_{-s}}{\lambda^{s}},\e

\n where $p_{1}=\pm \sqrt{q_2}$, i.e. there are two solutions of
(\ref{eq14}). Since these two solutions do not produce independent
integral densities due to the Lemma \ref{lem}, we proceed with, say,
$p_{1}=\sqrt{q_2}$.  The first nontrivial integral from here is

\b \label{eq18} h_{1}\equiv \int p_{1}\text{d}x= \int
\sqrt{q_2}\text{d}x. \e

\n From (\ref{eq14}) and (\ref{eq15}) we have

\b \label{eq19} 2p_{1}p_{0}+p_{1,x} =q_1, \qquad
p_{0}=\frac{q_1}{2\sqrt{q_2}}-\frac{q_{2,x}}{4q_2}.\e

\n Neglecting the exact derivative, we obtain the integral

\b \label{eqh0} h_{0} = \frac{1}{2}\int
\frac{q_1}{\sqrt{q_2}}\text{d}x. \e

\n The next equation,

\b \label{eq20} p_{0}^{2}+2p_{1}p_{-1}+p_{0,x}=\frac{1}{4},\e

\n gives

\b \label{eq20a} p_{-1}=\frac{1}{32}\Big(
\frac{4}{\sqrt{q_2}}+\frac{q_{2,x}^2}{q_2^{5/2}}-\frac{4q_1^2}{q_2^{3/2}}\Big)+\Big(\frac{q_{2,x}}{8q_{2}^{3/2}}-\frac{q_1}{4q_2}\Big)_x,\e

\n and thus we obtain the integral

\b \label{eq21} h_{-1}= \frac{1}{32}\int \Big(
\frac{4}{\sqrt{q_2}}+\frac{q_{2,x}^2}{q_2^{5/2}}-\frac{4q_1^2}{q_2^{3/2}}\Big)\text{d}x.
\e

It is not difficult to derive the general recurrent formula from
(\ref{eq14}) and (\ref{eq15}):

\b \label{eq22a} h_j&=&\int p_j\text{d}x,  \qquad
j=1,0,-1,...,\\\label{eq22} p_{1}&=&\sqrt{q_2}, \quad
p_{0}=\frac{q_1}{2\sqrt{q_2}}-\frac{q_{2,x}}{4q_2}, \quad
p_{-1}=\frac{1}{2p_{1}}\Big(\frac{1}{4}-p_{0}^{2}-p_{0,x}\Big),\\
 \label{eq23} p_{-j}&=&-\frac{1}{2p_{1}}\Big( \sum_{i=0}^{j-1}p_{-i}p_{-j+i+1}+p_{-j+1,x}\Big), \qquad j\geq 2. \e

 Now from (\ref{eq22}), (\ref{eq23}) and (\ref{eq50}) one can
 express the conservation laws in differential form (\ref{eqDF1}) with $a=1$:
\b \label{eq60} p_{1,t}&=& -(u_0 p_{1})_{x}, \nonumber\\
p_{0,t}&=&\Big(\frac{1}{2}u_{0,x}-u_0 p_0 - u_1 p_1\Big )_{x},
\nonumber \\
p_{-1,t}&=&\Big(\frac{1}{2}u_{1,x}+\frac{1}{2}p_{1}-u_1 p_0 - u_{0}
p_{-1}\Big )_{x},
\nonumber \\
p_{-j,t}&=&\Big(\frac{1}{2}p_{-j+2}-u_0p_{-j}-u_1p_{-j+1}\Big )_{x},
\qquad \qquad j\geq 2. \e

The reduction of (\ref{eq22a}) -- (\ref{eq60}) to the CH case
(Example 3) is straightforward via $q_1=u_1=0$. All even densities,
$p_0$, $p_{-2}$, ... are exact derivatives and do not produce
independent integrals of motion. Each of the remaining odd densities
$p_1$, $p_{-1}$, $p_{-3}$, ... produces only one independent
integral of motion. Thus the CH integrals are $h_1$, $h_{-1}$,
$h_{-3}$, $\ldots$ .

 Let us now take the expansion (\ref{eqB}). Since in this case
 $u_2=-1/2$, from (\ref{eqC}) for $r=k$ it follows that $p_{0,x}=0$ and then
 from (\ref{eq10}), $p_0=\pm 1/2$. Due to the Lemma \ref{lem}, we consider here only the first possibility,
 i.e. $p_0=1/2$:

 \b \label{eq24} p=\frac{1}{2}+\sum _{s=1}^{\infty}
p_{s}\lambda ^{s}.\e

\n Note that in (\ref{eq24}) $p_{1}$ is not the same as in
(\ref{eq15}). From (\ref{eq14}) and (\ref{eq24}) we obtain

\b \label{eq25} p_{1}+p_{1,x}=q_1, \e

\n which due to (\ref{L7}) has a solution

\b \label{eq26} p_{1} =u_1-u_{1,x}+\omega_1, \e

\n leading to the integral ($h_1$ is not the same as in
(\ref{eq18}))

\b \label{eq26a} h_{1} =\int u_1 \text{d}x. \e

\n The next equation from (\ref{eq14}) and (\ref{eq24}) is

\b \label{eq28} p_{2}+p_{2,x}+p_{1}^{2}=q_2. \e

\n Using (\ref{L8}), (\ref{eq26}) one can verify that

\b \label{eq29}
p_{2}=u_0-u_{0,x}+2u_1^2-2u_1u_{1,x}+2\omega_1u_1-\omega_1^2+\omega_2.
\e

\n Then the next independent integral is

\b \label{eq29a} h_{2}=\int(u_0+2u_1^2)\text{d}x+2\omega_1h_1. \e

\n The equation for $p_3$ is

\b \label{eq29b} p_{3}+p_{3,x}+2p_{1}p_2=0, \e

\n giving

\b \label{eq29c} h_{3}=\int p_3\text{d}x=-2\int p_1p_2\text{d}x, \e

\n where $p_1$ and $p_2$ are given by (\ref{eq26}), (\ref{eq29}). We
notice that formally (see (\ref{eq29b}))

\b \label{eq29d} p_{3}=-(1+\partial_x)^{-1}2p_{1}p_2. \e

\n The equation for $p_4$ is

\b \label{eq29e} p_{4}+p_{4,x}+2p_{1}p_3+p_2^2=0, \e

\n and thus

\b \label{eq29f} h_{4}=\int p_4\text{d}x=-\int
(2p_1p_3+p_2^2)\text{d}x. \e

\n In order to express $\int p_1p_3 \text{d}x$ only via the known
$p_1$ and $p_2$ we proceed as follows, using (\ref{eq29b}),
(\ref{eq26}) and neglecting total derivatives:

\b \label{eq29g}  &\phantom{*}&\int p_{1}p_{3}\text{d}x =-2\int
p_{1}^{2}p_2\text{d}x -\int p_{1}p_{3,x}\text{d}x=
\nonumber\\&\phantom{*}&-2\int p_{1}^{2}p_2\text{d}x -\int
(u_1-u_{1,x}+\omega_1)p_{3,x}\text{d}x=
\nonumber\\&\phantom{*}&-2\int p_{1}^{2}p_2\text{d}x -\int
(u_1p_{3,x}+u_{1,x}(2p_{1}p_{2}+p_{3}))\text{d}x=
\nonumber\\&\phantom{*}&-2\int p_{1}^{2}p_2\text{d}x -\int
((u_1p_{3})_x+2u_{1,x}p_{1}p_{2})\text{d}x=
\nonumber\\&\phantom{*}&-2\int
p_{1}p_2(u_{1,x}+p_{1})\text{d}x=-2\int
p_{1}p_2(u_1+\omega_1)\text{d}x= \nonumber\\&\phantom{*}&-2\int
p_{1}p_2u_1\text{d}x+\omega_1 h_3. \e

\n Finally,

\b \label{eq29h} h_{4}=\int
(4u_1p_1p_2-p_2^2)\text{d}x-2\omega_1h_3. \e

\n The densities for the higher integrals are, in general, nonlocal.

To summarize: the conserved quantities are

\b \label{eq36} h_{j}=\int p_{j}\text{d}x, \qquad j=1,2,\ldots, \e

\n where according to (\ref{eq14}) and (\ref{eq24}) the integral
densities $p_{j}$ can be computed recurrently:

\b \label{eq37} p_{j}=-(1+\partial_{x})^{-1}\sum _{i=1}^{j-1}
p_{i}p_{j-i}, \qquad j\geq 3, \e

\n where $p_{1}$ and $p_2$ are given in (\ref{eq26}), (\ref{eq29}),
e.g. see (\ref{eq29d}). Again from (\ref{eq37}) and (\ref{eq50}),
the conservation laws can be expressed in differential form
(\ref{eqDF2}):

\b p_{j,t}=\Big(\frac{1}{2}p_{j+2}-u_1p_{j+1}-u_0p_{j}\Big)_{x},
\qquad j=1,2,\ldots\phantom{*}. \e

The reduction to CH (Example 3)  goes as follows. With
$q_1=u_1=\omega_1=0$ we have (\ref{eq26}) $p_1=0$ and from
(\ref{eq29}), bearing in mind that in this case $u_0\equiv u$,
$\omega_2\equiv \omega$,

\b \label{eq38} p_{2}=u-u_x+\omega. \e

\n Then, clearly all odd integral densities are zero. Furthermore,
$h_2$ gives the CH integral $H_0=\int m\text{d}x$ and $h_4$ leads to
$H_1$ [recall  (\ref{eq2a})].  Now we can verify that the reduction
of $h_6$ leads to the second CH Hamiltonian $H_2$ [recall
(\ref{eq2b})].

The equation for $p_6$ with the reduction is

\b \label{eq39} p_6+p_{6,x}+2p_2p_4=0, \e

\n and correspondingly

\b \label{eq45} h_{6}=-2\int p_2p_4 \text{d} x. \e

\n The equation for $p_{4}$, (\ref{eq29e}) now is

\b \label{eq41} p_{4}+p_{4,x}+p_{2}^2=0, \e

\n and  therefore, in order to compute $\int p_{2}p_{4}\text{d}x$ we
proceed as follows. Multiplying (\ref{eq41}) by $p_{2}$ and then
using (\ref{eq41}) and (\ref{eq38}) we obtain:

\b  &\phantom{*}&\int p_{2}p_{4}\text{d}x =-\int p_{2}^{3}\text{d}x
-\int p_{2}p_{4,x}\text{d}x= \nonumber\\&\phantom{*}&-\int
p_{2}^{3}\text{d}x -\int (u-u_{x}+\omega)p_{4,x}\text{d}x=
\nonumber\\&\phantom{*}&-\int p_{2}^{3}\text{d}x -\int
(up_{4,x}+u_{x}(p_{2}^{2}+p_{4}))\text{d}x=
\nonumber\\&\phantom{*}&-\int p_{2}^{3}\text{d}x -\int
((up_{4})_x+u_{x}p_{2}^{2})\text{d}x= \nonumber\\&\phantom{*}&-\int
p_{2}^{2}(u_{x}+p_{2})\text{d}x=-\int p_{2}^{2}(u+\omega)\text{d}x=
\nonumber\\&\phantom{*}&-\int (u^{3}+uu_{x}^{2}+2\omega
u^{2})\text{d}x -3\omega ^{2} H_{0}-2\omega H_{1}, \nonumber \e

\n leading to the independent integral

\b  H_{2}=\frac{1}{2}\int (u^{3}+uu_{x}^{2}+2\omega u^{2})\text{d}x.
\nonumber \e

\n In other words, the CH conserved quantities are reproduced by the
even integrals $h_2$, $h_4$, $h_6$, $\ldots$ .

\section*{\it Acknowledgements}

The author is grateful to Prof. A. Constantin for helpful
discussions and to an anonymous referee for important suggestions.
The author also acknowledges funding from the Science Foundation
Ireland, Grant 04/BR6/M0042.

\end{document}